\documentclass[12pt]{article}

\usepackage{scicite}

\usepackage{times}
\usepackage{amsmath,amstext,amsthm,amssymb,amsxtra}

\usepackage{graphicx}

\newcommand{\RR}{{\mathbb{R}}}
\newcommand{\NN}{{\mathbb{N}}}

\newcommand{\eps}{\varepsilon}
\newcommand{\bp}{\noindent {\it Proof}.\,\,}
\newcommand{\ep}{\hfill$\Box$ \vskip 0.08in}

\newcommand{\po}{{\partial\Omega}}

\newtheorem{proposition}{Proposition}[section]

\def\ring{\mathaccent"0017 }

\newenvironment{sciabstract}{%
\begin{quote} \bf}
{\end{quote}}

\newcounter{lastnote}
\newenvironment{scilastnote}{%
\setcounter{lastnote}{\value{enumiv}}%
\addtocounter{lastnote}{+1}%
\begin{list}%
{\arabic{lastnote}.}
{\setlength{\leftmargin}{.22in}}
{\setlength{\labelsep}{.5em}}}
{\end{list}}

\title{The Hidden Landscape of Localization}

\author
{Marcel Filoche$^{1,2\ast}$, Svitlana Mayboroda$^{3}$\\
\\
\normalsize{$^{1}$Physique de la Mati\`ere Condens\'ee, Ecole Polytechnique, CNRS,}\\
\normalsize{91128 Palaiseau, France}\\
\normalsize{$^{2}$CMLA, ENS Cachan, CNRS, UniverSud,} \\
\normalsize{61 Avenue du Pr\'esident Wilson, F-94230 Cachan, France}\\
\normalsize{$^{3}$Department of Mathematics, Purdue University,}\\
\normalsize{150 North University Street, West Lafayette, Indiana 47907-2067, USA}\\
\\
\normalsize{$^\ast$Marcel Filoche; E-mail:  marcel.filoche@polytechnique.edu.}
}

\date{}

\begin{document} 


\baselineskip24pt


\maketitle


\begin{sciabstract}
Wave localization occurs in all types of vibrating systems, in acoustics, mechanics, optics, or quantum physics. It arises either in systems of irregular geometry (weak localization) or in disordered systems (Anderson localization). We present here a general theory that explains how the system geometry and the wave operator interplay to give rise to a ``landscape" that splits the system into weakly coupled subregions, and how these regions shape the spatial distribution of the vibrational eigenmodes. This theory holds in any dimension, for any domain shape, and for all operators deriving from an energy form. It encompasses both weak and Anderson localizations in the same mathematical frame and shows, in particular, that Anderson localization can be understood as a special case of weak localization in a very rough landscape.
\end{sciabstract}


\paragraph*{Introduction}\label{Intro}

Every object in nature vibrates. The vibrations can be acoustic waves in a medium, mechanical deformations of a rigid plate, electromagnetic waves in a cavity or quantum states such as the electronic states in a crystal. In all these cases, the linear behavior of the system can be reduced to the knowledge of the vibrational modes, i.e., the eigenfunctions and eigenvalues of the spatial differential operator associated to the wave equation. A stunning property exhibited by these eigenfunctions in irregular or disordered systems is known as localization: in some cases, the mode amplitude is small everywhere except in a very limited subregion of the domain, even though there is no clearly visible obstacle preventing the vibration to propagate in the rest of the domain, away from its main existence subregion.

It has been acknowledged that there are several types of localization, each exhibiting a specific behavior in space or in frequency. First, when caused by the irregular or complex geometry of the vibrating domain, localization is classified as \emph{weak}. It is characterized by a slow decay of the mode amplitude away from its main existence subregion (much slower than exponential) \cite{Baranger1993,Felix2007,Filoche2009,Heilman2010}. Secondly, localization can arise due to a disorder quenched in the system~\cite{Anderson1958,Richardella2010}. This phenomenon, called \emph{Anderson localization}, has fascinated scientists since its discovery in 1958 and spurred a wealth of literature~\cite{Lagendijk2009}. In that case, the mode amplitude decays exponentially away from its main existence subregion and the localization is classified as \emph{strong}. Finally, a third and quite different type of localization occurs at high frequencies in specific domains that possess stable orbits such as whispering gallery or bouncing ball modes~\cite{Heller1984}. Confined modes known as \emph{scar} modes then appear asymptotically localized near these orbits.

\begin{figure}
\center
\includegraphics[width=4.25cm]{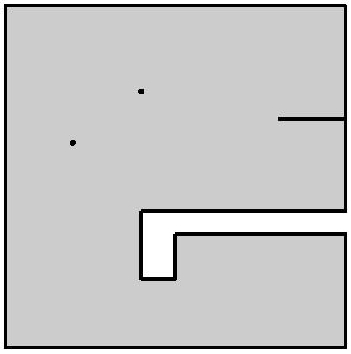} \hskip 5mm
\includegraphics[width=5.4cm]{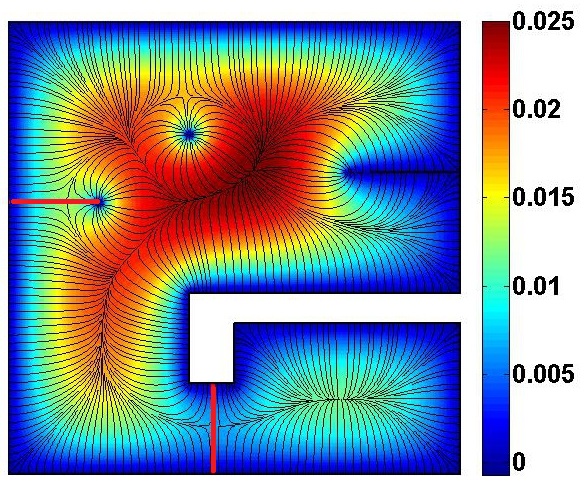} \hskip 2mm
\includegraphics[width=5.4cm]{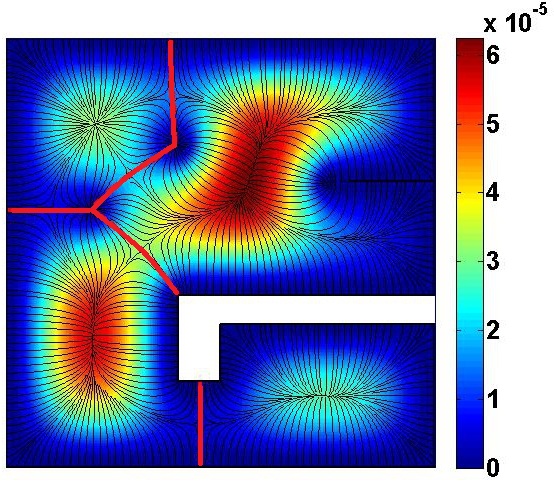}
\caption{Left: Geometry of a complex domain, with a bottleneck (in the lower part), two inner blocked points, and an inward crack (on the right upper side). The question is: is there localization in this structure, which modes will be localized and where? Center and right: 2D representations of the landscape $u$ for the Laplacian (center) and the bi-Laplacian (right) in the same domain. The colors correspond to the height of the landscape. The (thin black) streamlines  help to detect the valley lines, which are then highlighted as thick red curves. These valleys delimit 2 subregions of localization for the Laplacian and 4 subregions for the bi-Laplacian.}
\label{fig:cplate_u}
\end{figure}

Until today, there has been no theory able to explain how the geometry of the domain or the nature of the disorder is related to the localization of vibrations, to predict in which subregions one can expect localized eigenmodes to appear, and in which frequency range. Moreover, the question of whether the different types of localization are linked remains open. Consider, for instance, the system depicted in Fig.~\ref{fig:cplate_u} (left). It has a non trivial shape, possesses 2 inner blocked points (in the left upper region), one crack on the right upper boundary and a bottleneck in its lower region. This can represent either a flexible membrane of complex shape (the spatial differential operator is then the Laplacian), or a rigid thin plate (the operator being the bi-Laplacian). From the knowledge of this geometry only, how can one determine whether and where to expect localization in this structure? In the present paper, we show that one theory can answer these questions. It unifies weak and Anderson localization, and reveals inside each system a hidden landscape that determines the localization subregions, the strength of the confinement, and its frequency dependency.

\paragraph*{The landscape mapping} \label{landscape}

In mathematical terms, a vibrating system is governed by the wave equation associated to a suitable elliptic differential operator $L$. The latter is determined by the nature of vibration and the medium. For instance,  the Laplacian $L=-\Delta$ describes the vibration of a soft membrane in 2D, acoustic waves, the quantum states inside a box or cavity; variable coefficient second order operators $L=-{\rm div} \left( A(x)\,\nabla \right)$ pertain to the aforementioned phenomena in inhomogeneous media; and the bi-Laplacian $\Delta^2$ addresses thin plate vibrations in 2D. A study of the vibrational properties of the system can be reduced to the investigation of the eigenmodes of $L$ defined by 
\begin{equation}\label{eq11.0}
L~\varphi~=~\lambda~\varphi\,\,\mbox{ in } \,\,\Omega,\quad \varphi|_{\partial\Omega}=0
\end{equation}
We demonstrate here that {\it all} eigenmodes are controlled by the same function which has a decisive impact on localization properties. To be specific, for every $\lambda$ and $\varphi$ as above
\begin{equation}\label{eq9}
|\varphi(\vec{x})|\leq \lambda \int_\Omega |G(\vec{x},\vec{y})|\,dy,\quad \forall\,\vec{x}\in\Omega,
\end{equation} 
\noindent where the mode $\varphi$ is normalized so that $\sup_\Omega |\varphi|=1$, and $G(\vec{x},\vec{y})$ is the Green's function solving $L~G(\vec{x},\vec{y})=\delta_{\vec{x}}(\vec{y})$ with zero data on the boundary. (For the sake of brevity, we assume Dirichlet boundary data. Other types of boundary conditions will be addressed in forthcoming publications). In other words, there exists a function $u(\vec{x})=\int_\Omega |G(\vec{x},\vec{y})|\,dy$, \emph{ independent} of the eigenmode $\varphi$ such that 
\begin{equation}\label{eq10}
{|\varphi(\vec{x})|}\leq \lambda \,u(\vec{x})
\end{equation} 
\noindent Recall that the Green's function is positive for the Laplacian and, more generally, for all second order differential operators. In that case $u$ admits a remarkably simple definition. It is the solution to the Dirichlet problem 
\begin{equation}\label{eq11}
L~u~=~1 \,\,\mbox{ in } \,\,\Omega,\quad u|_{\partial\Omega}=0
\end{equation}
\noindent For a higher order operator the boundary condition $u|_{\partial\Omega}=0$ in \eqref{eq11.0} and \eqref{eq11} refers to the natural Dirichlet data, including vanishing derivatives of appropriate orders. We defer rigorous mathematical justification of \eqref{eq9}--\eqref{eq11} to~\cite{supporting}. In physical terms, $u$ can be interpreted, for instance, as the steady-state deformation of a membrane under a uniform load. 

Through inequality \eqref{eq10}, the \emph{landscape} $u$ compels the eigenmodes to be small along its lines of local minima (called {\it valleys} throughout the paper). As we will show, the network of these valleys, a priori invisible when looking at the domain, but clearly identifiable on the graph of $u$, operates as a driving force that determines the confinement properties for both weak and Anderson localization. 

Let us illustrate this with an example. Figure~\ref{fig:cplate_u} displays two maps of $u$ computed in the same complex geometry for the Laplacian (center) and the bi-Laplacian (right), respectively. The streamlines (lines of the gradient) have been plotted in thin black in order to clearly pinpoint the valleys, which are highlighted as thick red lines. The two cases expose dramatically different patterns. For the Laplacian, one can observe two valleys, and only one of them splits the domain into two disjoint subregions. In the bi-Laplacian case, five valleys form a network  yielding a partition of the domain into four disjoint subregions.

We now show rigorously that for any domain and any differential operator the network of valleys built as above triggers wave localization by  effectively separating the original domain into weakly coupled vibrating regions.

\paragraph*{The formation of localized modes}\label{formation}

Consider a subregion $\Omega_1$ of the original domain $\Omega$ carved out by the valley lines. It can be thought of, for instance, as any of the four subregions in Figure~\ref{fig:cplate_u}, right. By construction, $u$ is relatively small (locally minimal) along the boundary of $\Omega_1$. Thus, for low eigenvalues $\lambda$, inequality \eqref{eq10} provides a severe constraint on the mode amplitude $\varphi$ along the boundary $\partial\Omega_1$. As a result, any eigenmode $\varphi$ of the entire domain can locally be viewed as a solution to the problem
\begin{align}\label{eqED1}
L~\varphi &= \lambda~\varphi \qquad \rm{in} \quad \Omega_1,\\
\varphi = 0 \qquad \rm{on} \quad \partial \Omega_1\cap \partial\Omega , \qquad & \rm{and} \qquad \varphi = \eps \qquad \rm{on} \quad \partial \Omega_1\setminus \partial\Omega,\label{eqED1.0}
\end{align}
\noindent where $\eps(\vec{x})$ is a quantity smaller than $\lambda u(\vec{x})$ on the boundary of $\Omega_1$. Observe that boundary value problem \eqref{eqED1}--\eqref{eqED1.0} is, in fact, akin to the eigenvalue problem in the subregion $\Omega_1$ alone: the differential equation inside the subregion is identical, but on the boundary $\varphi$ is small in \eqref{eqED1.0} rather than just being zero as an eigenvalue problem on $\Omega_1$ would normally warrant.

Using properties of the resolvent $(L-\lambda I)^{-1}$ and spectral decomposition, one can establish the following estimate which plays an essential role in understanding the origin of localization:
\begin{equation}\label{eq19}
\|\varphi\|_{L^2(\Omega_1)}  \leq  \left(1+\frac{\lambda}{d_{\Omega_1}(\lambda)}\right) \|\eps\|
\end{equation}
\noindent See~\cite{supporting} for the proof. Here, $d_{\Omega_1}(\lambda)$ is the distance from $\lambda$ to the spectrum of the operator $L$ in the subregion $\Omega_1$ (defined as: $\displaystyle d_{\Omega_1}(\lambda) = \min_{\lambda_{k,\Omega_1}} \left\{\left|\lambda-\lambda_{k,\Omega_1}\right|\right\}$, the minimum being taken over all eigenvalues $\left(\lambda_{k,\Omega_1}\right)$ of $L$ in $\Omega_1$), and $\|\eps\|$ is the $L^2$-norm of the solution to $Lv=0$ in $\Omega_1$ with data $\eps$ on $\partial\Omega_1$ (in the sense of \eqref{eqED1.0}). In particular, $\|\eps\|$ becomes arbitrarily small as $\eps$ in Eq.~\eqref{eqED1.0} vanishes.

The presence of $d_{\Omega_1}(\lambda)$ in the denominator of the right-hand side of Eq.~\eqref{eq19} assures that whenever $\lambda$ is {\it far} from any eigenvalue of $L$ in $\Omega_1$ in relative value, the norm of $\varphi$ in the entire subregion, $\|\varphi\|_{L^2(\Omega_1)}$, has to be smaller than $2\|\eps\|$. Consequently, such a mode $\varphi$ is expelled from $\Omega_1$ and must ``live" in its complement, exhibiting weak localization. Conversely, $\varphi$ can only be substantial in the subregion $\Omega_1$  when $\lambda$ almost coincides with one of local eigenvalues of the operator $L$ in $\Omega_1$. Moreover, in that case Eq.~\eqref{eqED1} yields the conclusion that $\varphi$ itself almost coincides with the corresponding eigenmode of the subregion $\Omega_1$.

Thus, we obtain a rigorous scheme elucidating the formation of weak localization.
In any subregion delimited by the valleys of $u$, an eigenmode of $\Omega$ has only two possible choices: (1) either its amplitude is very small throughout this subregion, or (2) this mode mimics (both in frequency and in shape) one of the subregion's own eigenmodes. Consequently, a low frequency eigenmode can cross the boundary between two adjacent subregions only if they possess two similar local eigenvalues. More generally, a {\it fully delocalized} eigenmode can only emerge as a {\it collection} of local eigenmodes of all subdomains when they all share a common eigenvalue.

\begin{figure}
\center
\includegraphics[width=3.6cm]{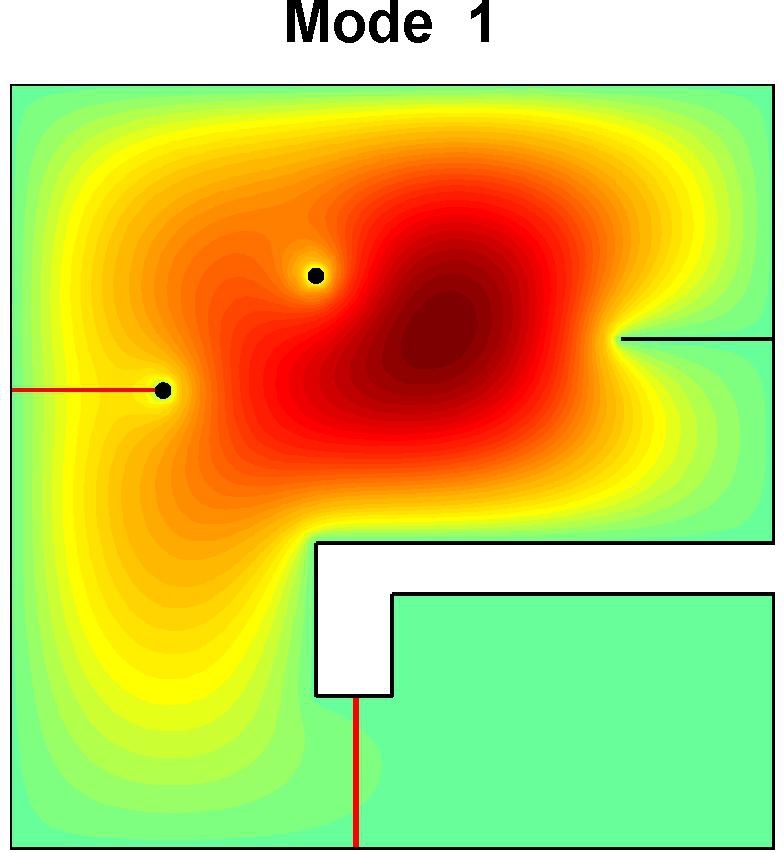}\hskip 2cm
\includegraphics[width=3.6cm]{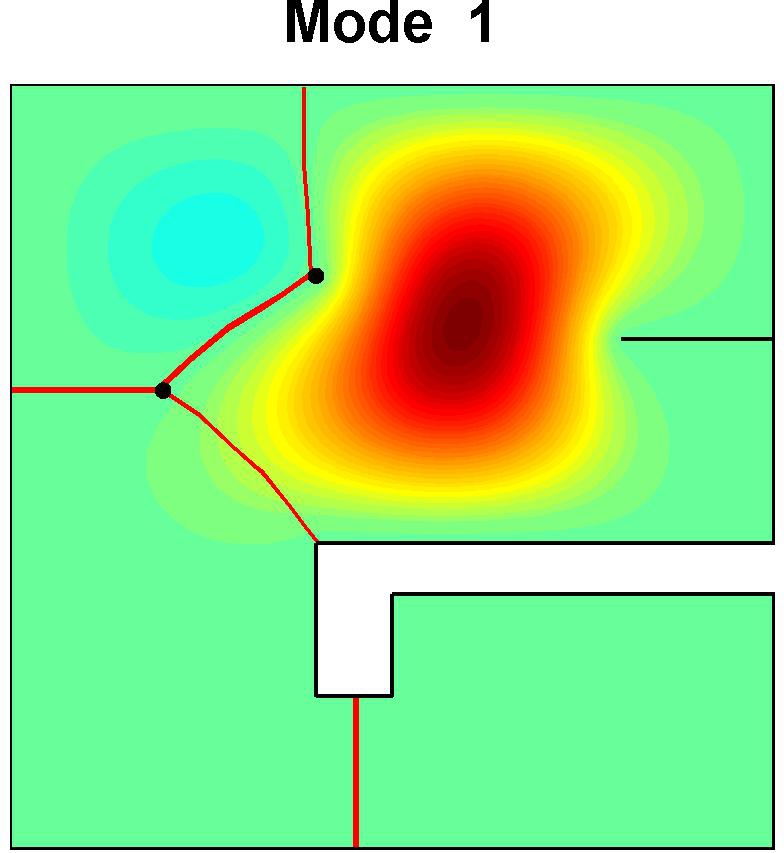}\hskip 2mm
\includegraphics[width=3.6cm]{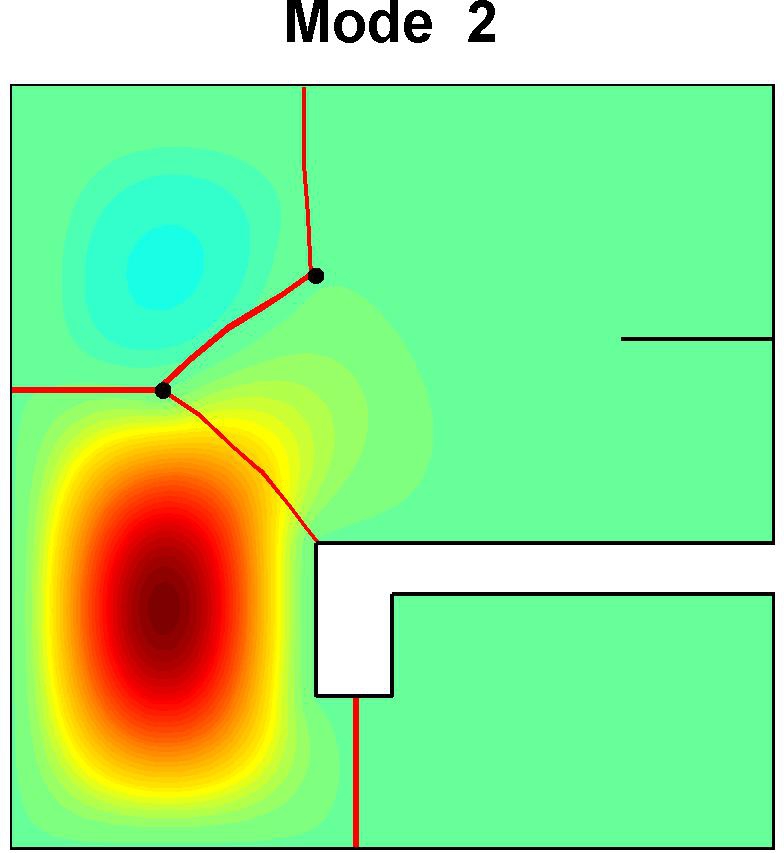}\\\vskip 2mm
\includegraphics[width=3.6cm]{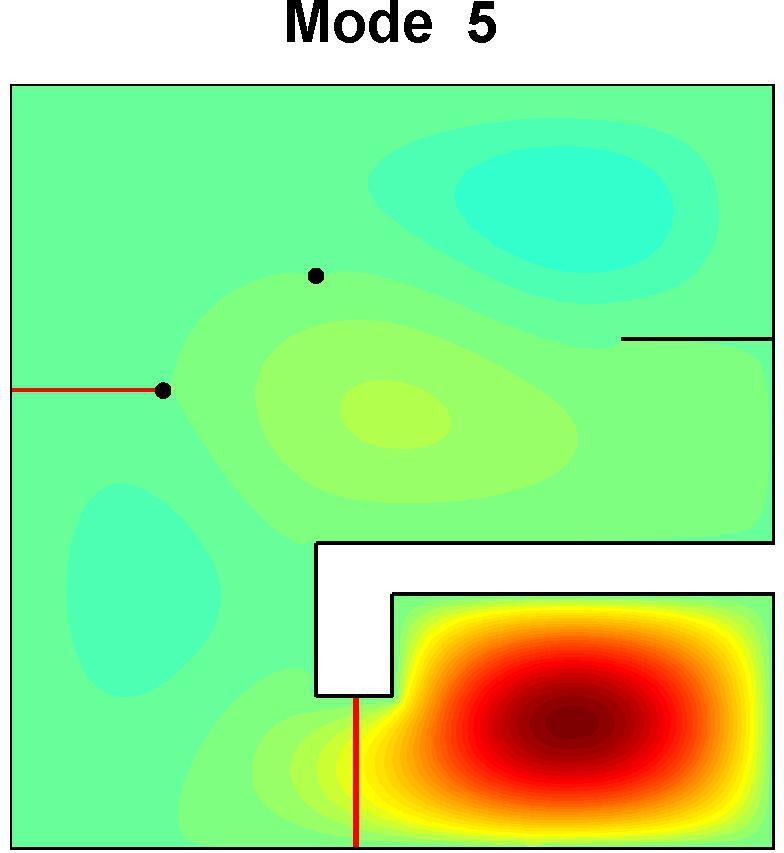}\hskip 2cm
\includegraphics[width=3.6cm]{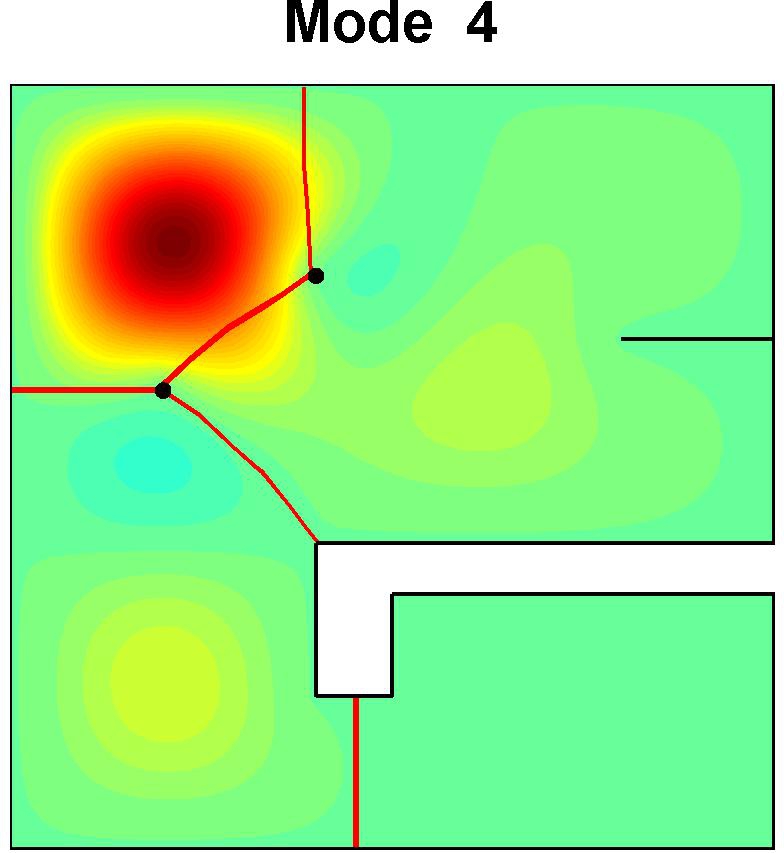}\hskip 2mm
\includegraphics[width=3.6cm]{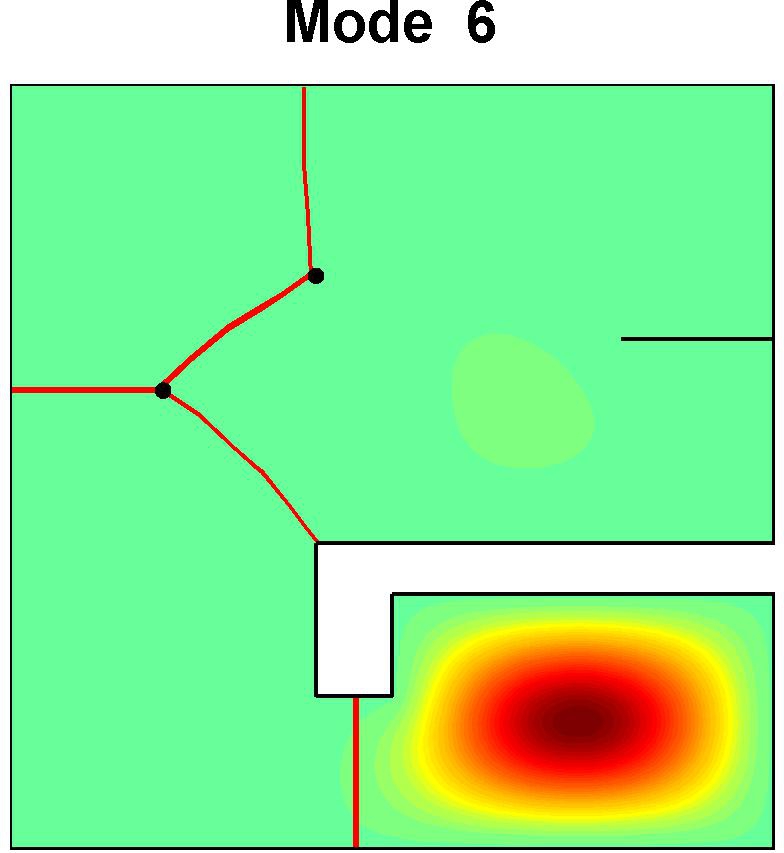}
\caption{two localized modes of the Laplace operator (left) and four localized modes of the bi-Laplace operator (right). In both cases, the valley network (displayed in red) obtained in Figure~\ref{fig:cplate_u} accurately predicts the number and the location of the localization subregions.}
\label{fig:cplate_modes}
\end{figure}

The role played by the valleys of $u$ in localization is even more astonishing when exploring the examples. Let us revisit the domain in Figure~\ref{fig:cplate_u}. Figure~\ref{fig:cplate_modes}, left, displays two localized modes (number 1 and 5) of the Laplacian plotted together with the valley lines computed in Figure~\ref{fig:cplate_u}, center. Not only these modes obey the pattern predicted by the landscape $u$, but there exists no eigenmode confined to a smaller subregion. In Figure~\ref{fig:cplate_modes}, right, modes 1, 2, 4, and 6 of the bi-Laplacian are displayed in the same domain together with the valley lines from Figure~\ref{fig:cplate_u}, right. We observe here a very different and much larger variety of localization behaviors. However, once again, the location, the shape, and the number of the localization subregions exactly match the partition of the domain generated by the valley network of the landscape of  $u$ computed in Figure~\ref{fig:cplate_u}.

\paragraph*{The effective valley network and Weyl's law}\label{effective}

Not only the location of the subregions is revealed by the network of valleys, but the height and shape of the valleys gives access to statistical information on the evolution of the localization properties in {\it all} frequency ranges. While geometrically the network of valleys is determined by $u$ and does not depend on a particular eigenmode, the {\it strength of the confinement} of an eigenmode dictated by Eq.~\eqref{eq10} diminishes as $\lambda$ grows. Indeed, given the normalization chosen in Eq.~\eqref{eq9}, Eq.~\eqref{eq10} represents an effective constraint only at those points $\{\vec{x}\}$ where $\lambda~u(\vec{x})$ is smaller than 1. In other words, an eigenmode can actually  {\it see} only the {\it portion} of the initial landscape which satisfies $u<1/\lambda$. We refer to the latter as the \emph{effective} valley network.

By definition, at relatively low frequencies the effective network is identical to the full one. However, as $\lambda$ increases, the effective network progressively disappears. Subregions that were initially disjoint begin to merge to form larger subregions. Above a critical value of $\lambda$, what is left of the effective network allows one subregion to percolate throughout the system: starting from that value of $\lambda$, completely new fully delocalized modes can appear. 

Note that the precise quantitative information on evolution of the effective network with the growth of $\lambda$ is already encoded in the original mapping of $u$. In this vein, it would be extremely useful to estimate the underlying rate of growth of the eigenvalues. Unfortunately, in the full generality of our set-up the Weyl's law is not available yet. Roughly speaking, it predicts that for any given $\Lambda>0$ the number of the eigenvalues of $L$ below $\Lambda$ is asymptotically $\displaystyle \Lambda^{d/2m}$, where $d$ stands for the dimension and $2m$ is the order of the differential operator $L$. This estimate is widely used in practice and can be employed in the present context to determine the range of frequencies with high response to the impact of $u$, that is, the range of thoroughly localized eigenmodes. In this context, one has to stress that in a domain of self-similar boundary, smaller copies of a valley line would appear at all scales, triggering mode localization for an infinite number of eigenvalues. In that case, even though weak localization essentially affects low frequency eigenmodes, one would always find localized modes in the high frequency limit governed by Weyl's law.

\paragraph*{From weak to strong: Anderson localization}\label{anderson}

Since Anderson's work in 1958~\cite{Anderson1958}, strong localization is known to arise due to the presence of structural disorder in a system. For instance, in a gas of non interacting electrons in a crystal, this phenomenon can induce a transition between metallic and insulating behavior as the electronic states become highly localized~\cite{Punnoose2005,Richardella2010}.
More recently, disorder-induced localization has been demonstrated to be a very general phenomenon also observed in acoustics~\cite{Zhang1999}, in microwaves~\cite{Laurent2007}, or in optics~\cite{Sapienza2010,Riboli2011}.

\begin{figure}
\center
\includegraphics[width=7cm]{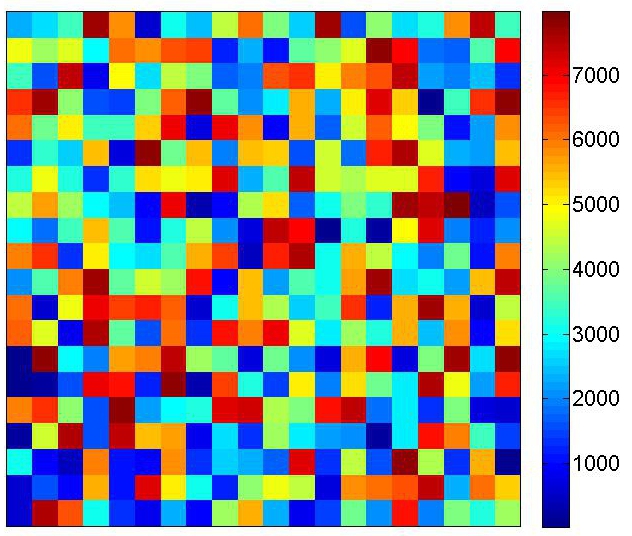}\hskip 5mm
\includegraphics[width=7cm]{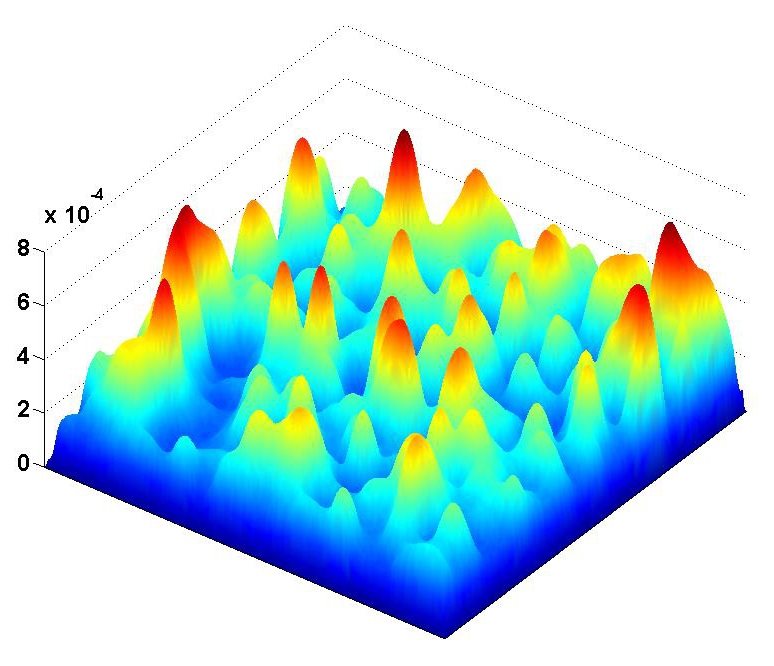} \vskip 5mm
\includegraphics[width=8cm]{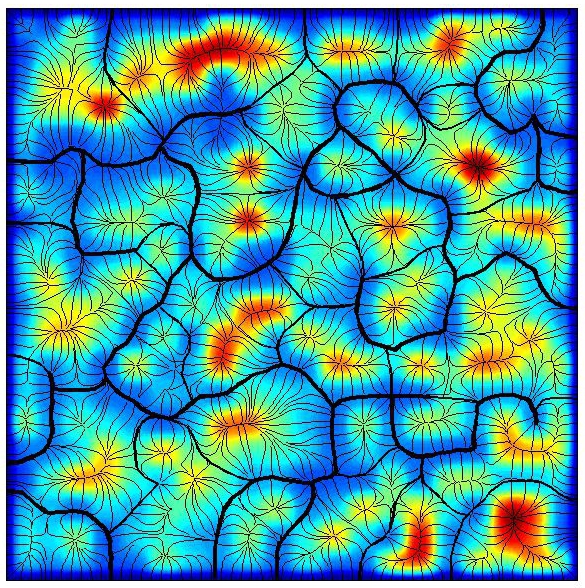}
\caption{Top left: random potential $V(\vec{x})$. The 2D square domain is divided in 20 $\times$ 20 small squares. In each square, the potential $V$ is assigned an independent random value uniformly distributed between 0 and $V_{max}$ (here 8000). Top right: 3D view of the landscape of $u$, obtained by solving $L u = 1$, where $L = - \Delta + V(\vec{x})$. Bottom: 2D color representation of the map of $u$, together with the streamlines. The thicker lines correspond to the deepest valleys and the moderately thick lines to the more shallow ones. This landscape draws an intricate network of interconnected valleys. One can conjecture that in the limit of a Brownian potential, this network becomes scale invariant and exhibit fractal properties.}
\label{fig:random_u}
\end{figure}

We present here a \emph{fundamentally novel approach} to Anderson localization, resting upon the theory developed in the previous sections.
To this end, we compute eigenmodes of the Schr\"{o}dinger operator with a random potential modeled as follows. The original domain is chosen to be a simple unit square. It is divided into $400=20\times 20$ smaller squares. On each of these smaller squares, the potential $V$ is constant, its value being determined at random uniformly between 0 and $V_{\max}$ (here $V_{\max}=8000$, see Figure~\ref{fig:random_u}, top left, the arbitrary energy units are taken such that $\hbar^2/2m=1$). The corresponding Schr\"{o}dinger operator $H = -\Delta + V$  is therefore a second order elliptic operator with variable coefficients falling under the scope of our theory. The valley landscape is further obtained by solving $H~u = 1$ in the entire square, with Dirichlet boundary conditions on 4 sides (see Figure~\ref{fig:random_u}, top right). One can observe a complex relief, characterized by a network of interconnected valleys of varying depths. Using the streamlines as guidelines, the valley lines are drawn in Figure~\ref{fig:random_u}, bottom. The thick lines represent the deepest valleys, while the moderately thick lines correspond to the shallower ones. This intricate network reveals a complex partition of the domain into a large number of subregions that was impossible to guess by just looking at the random potential at hand (cf. Figure~\ref{fig:random_u}, top left). In the continuous limit of a Gaussian random potential, one can conjecture that this network would show similar patterns at all possible valley depths, and hence, exhibit fractal properties.

\begin{figure}
\center
\includegraphics[width=3.3cm]{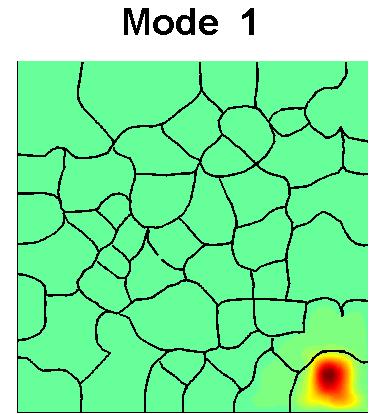}\hskip 3mm
\includegraphics[width=3.3cm]{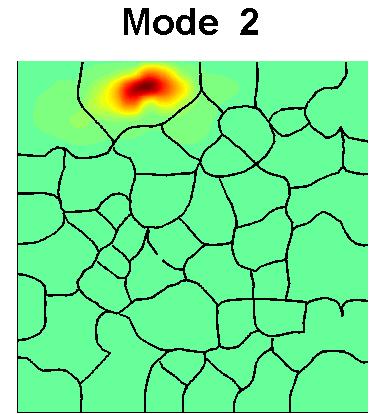}\hskip 3mm
\includegraphics[width=3.3cm]{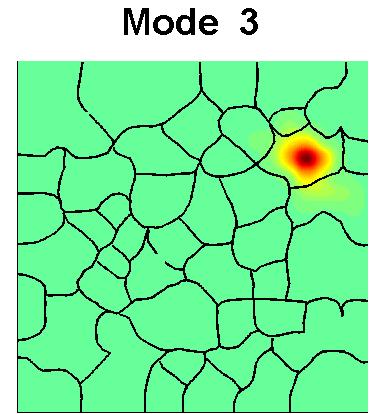}\hskip 3mm
\includegraphics[width=3.3cm]{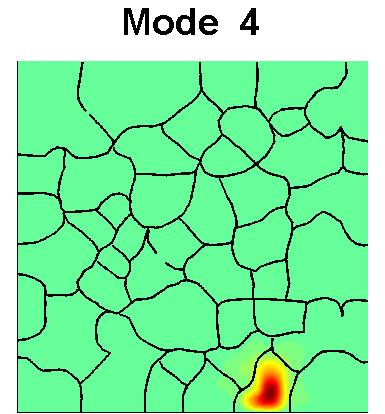}\vskip 2mm
\includegraphics[width=3.3cm]{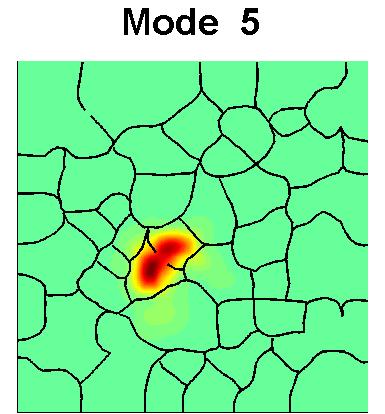}\hskip 3mm
\includegraphics[width=3.3cm]{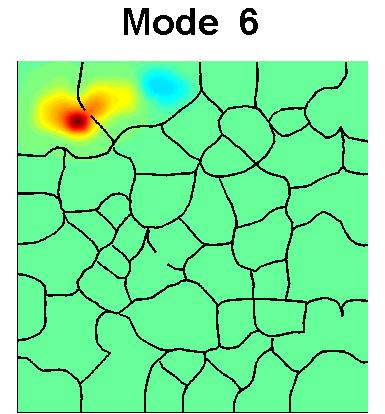}\hskip 3mm
\includegraphics[width=3.3cm]{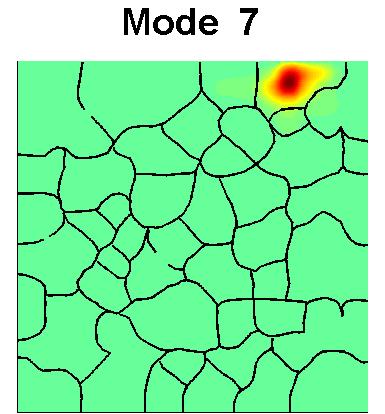}\hskip 3mm
\includegraphics[width=3.3cm]{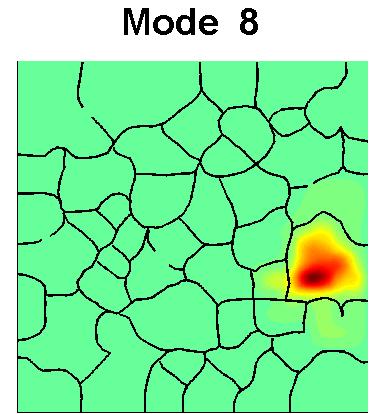}\vskip 2mm
\includegraphics[width=3.3cm]{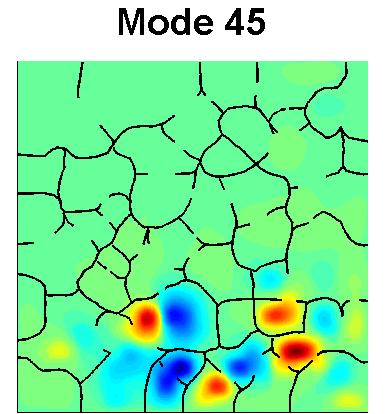}\hskip 3mm
\includegraphics[width=3.3cm]{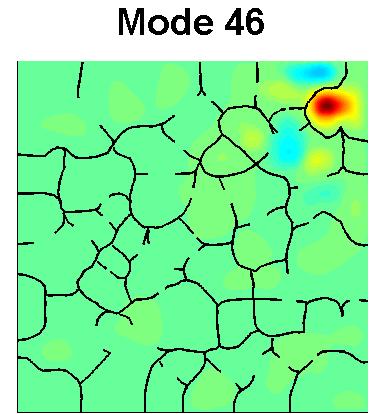}\hskip 3mm
\includegraphics[width=3.3cm]{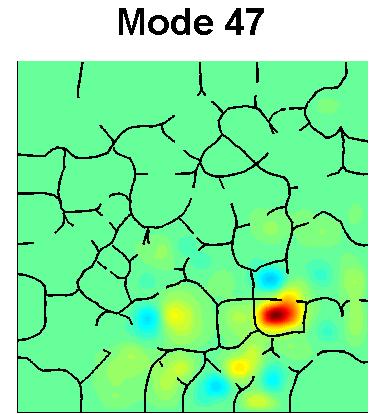}\hskip 3mm
\includegraphics[width=3.3cm]{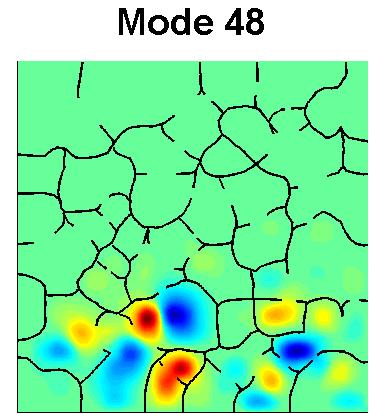}\vskip 2mm
\includegraphics[width=3.3cm]{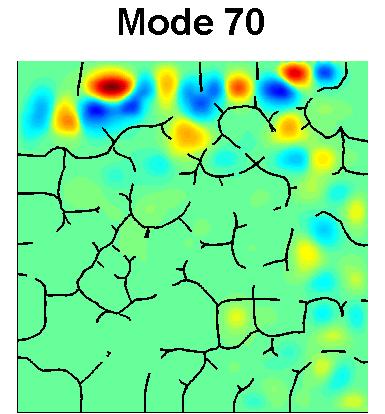}\hskip 3mm
\includegraphics[width=3.3cm]{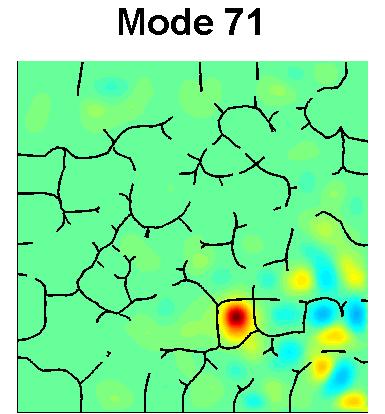}\hskip 3mm
\includegraphics[width=3.3cm]{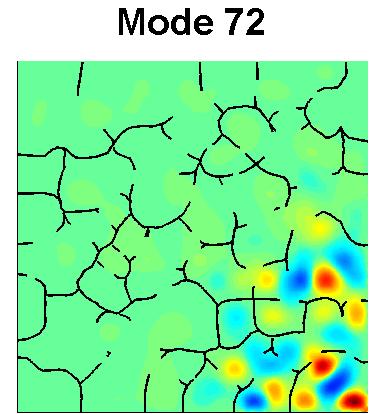}\hskip 3mm
\includegraphics[width=3.3cm]{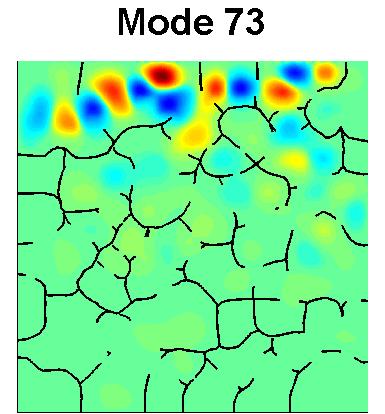}\vskip 2mm
\includegraphics[width=3.3cm]{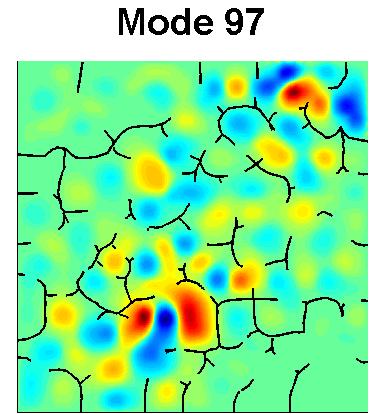}\hskip 3mm
\includegraphics[width=3.3cm]{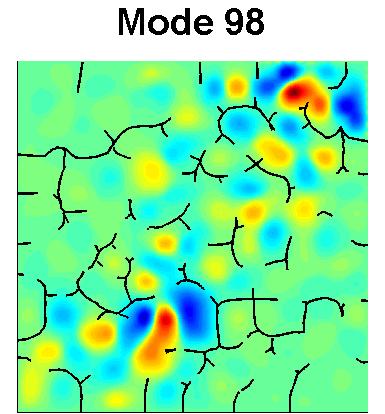}\hskip 3mm
\includegraphics[width=3.3cm]{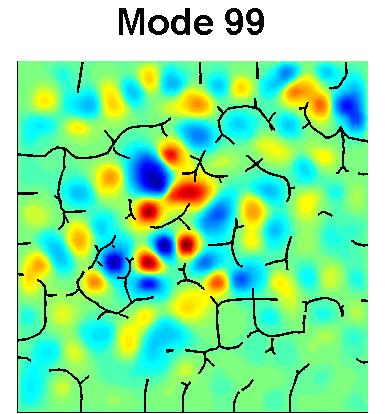}\hskip 3mm
\includegraphics[width=3.3cm]{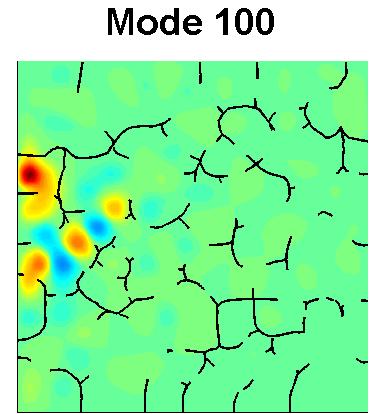}
\caption{Spatial distribution of several quantum states in the random potential of Fig.~\ref{fig:random_u}. On top of each state is drawn the effective valley network corresponding to the state energy. One can clearly observe that all modes are localized exactly in one of the subregions delimited by the effective valley network.}
\label{fig:random_modes}
\end{figure}

Figure~\ref{fig:random_modes} displays 2D color representations of the amplitude for a number of modes (or quantum states), at lower and higher frequencies. For each mode, the \emph{effective} network of valley lines is plotted on top of the amplitude. Since in the present context the eigenvalue is equal to the energy, the effective network can be viewed as a subset of the initial landscape of valleys subject to constraint  $u< 1/E$, $E$ being the mode energy. It is striking to observe how all modes are clearly shaped by the valley lines. The fundamental and the first excited states (modes 1 to 8) are localized completely to one of the subregions defined by the network of valley lines. At higher energies (mode 45, 48, 70), the effective valley network starts to shrink, opening breaches in the shallowest valley lines. One can see that these modes are still localized, but now exactly in the much larger subregions defined by the remaining effective network. However, there still exist some small subregions in which one can find localized modes weakly coupled to the rest of the domain (modes 47 and 71). At even higher energies (modes 97-99), the effective valley network is mostly disconnected, allowing a subregion to percolate throughout the entire domain: delocalized states appear.

For sake of simplicity, simulations have been carried out in 2D. In 3D, the valleys are not lines but surfaces of minimal value of $u$. The entire valley network has the shape of a foam which separates the domain into a large number of subregions. When the energy increases, the effective valley network evolves by opening gaps in the walls separating adjacent subregions. The localization length then increases accordingly as the average size of a subregion.

\begin{figure}
\center
\includegraphics[width=5cm]{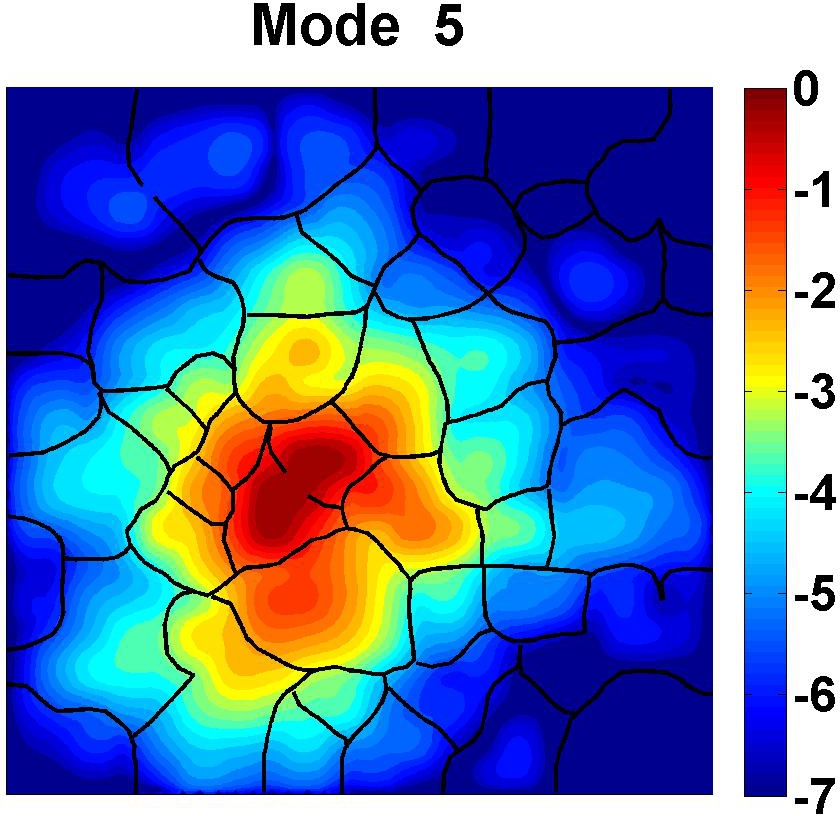}\hskip 2cm
\includegraphics[width=5cm]{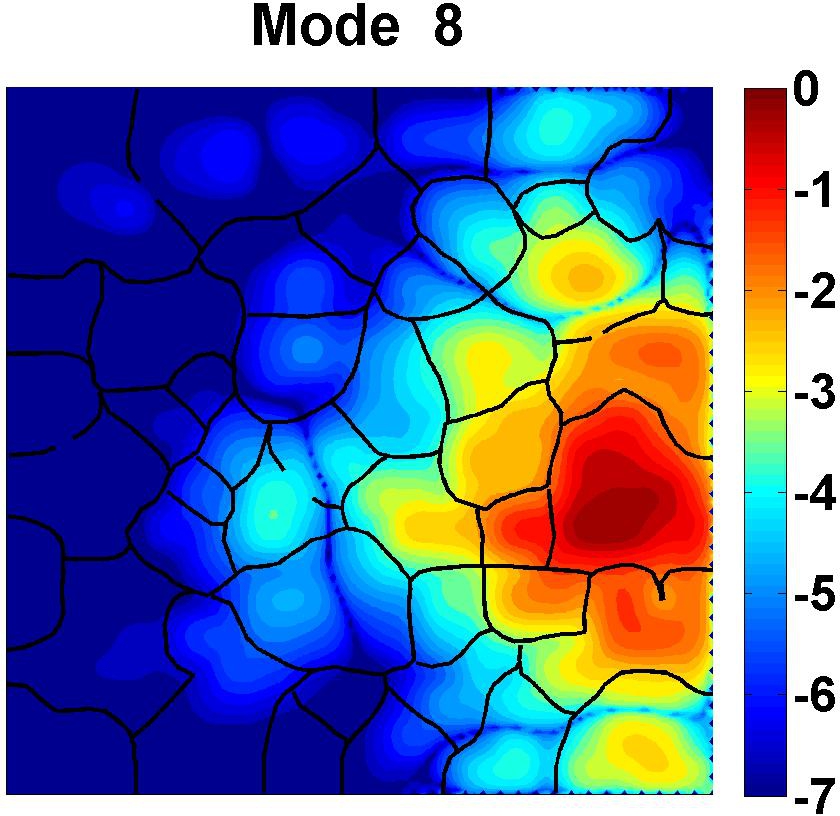}
\caption{Logarithmic plot (in log10) of the amplitude for two excited states
in the potential shown in Fig.~\ref{fig:random_u}. One can observe here in detail how strong localization emerges from weak localization. Both modes are located in one of the subregions delimited by the valleys ways. Moreover, their decay is shaped by neighboring subregions. Firstly, the mode amplitude is more or less uniform within any subregion. Secondly, when going away from the main subregion, crossing a valley line corresponds to a decrease by one or two orders of magnitude. This leads to an exponential decay for distance larger than the typical size a subregion.}
\label{fig:log_modes}
\end{figure}

One can examine in detail the strength of the localization by plotting the mode amplitude on a logarithmic scale (Figure~\ref{fig:log_modes}). By doing so, one can notice that the level curves are on average equally spaced, which corresponds to an exponential decay away from the existence subregion. Even more precisely, we once again observe our landscape at work: the amplitude of the mode appears to stay within the same order of magnitude inside each subregion, the decay essentially occurs when crossing a boundary between two adjacent subregions (this boundary corresponds to a valley line). This is particularly clear for both modes in Figure~\ref{fig:log_modes} in which their principal existence subregion appears dark red, all nearest neighboring subregions appear light red, the second nearest neighboring subregions essentially orange, etc. Therefore, each subregion is weakly coupled to its neighbors, the mode decaying by a somewhat constant factor each time it crosses a valley line away from the center subregion of existence. Hence, when zoomed towards any subregion, mode localization is of the weak type. However, due to the intricacy of the valley network, the succession of regularly spaced valley lines in the effective network yields an exponential decay of the mode away from its center subregion, over distances much larger than the typical size of a subregion. As a consequence, the mode appears to be exponentially confined: strong localization emerges from successive weak localizations.

Therefore, the exact same concepts and the exact same theory both account for localization in a domain of irregular geometry and localization induced by the presence of disorder. Moreover the notion of effective network extremely accurately predicts the shape of the localization subregions at any energy and therefore allows us to understand the apparition of delocalized states above a critical energy: this critical value is the smallest energy for which the effective network opens all valley lines and allows one subregion to percolate throughout the system. Remarkably, computing the landscape of $u$ and the effective valley network at any energy does not require any knowledge on the quantum states of the system, but yet gives access to accurate information on the confinement of the states within any energy range. One also should add that, although the landscape $u$ is obtained by simply solving one linear system, it nevertheless depends in a complex way upon the quenched disorder introduced in the potential~$V$. In particular, different types of randomness would lead to different localization properties. More generally, the question of localization in any disordered system can now be mathematically reformulated in the simple following way: How does the valley network of the landscape of $u$ (its connectivity, sizes, and height), depend on the properties of the disorder in $V$?

Localization also directly affects the transport properties in a random medium~\cite{Wang2011}: at zero temperature, the system is insulating if all the quantum states occupied by the electrons are localized. This condition can be reformulated in terms of the landscape: {\it for the system to be insulating at T=0, its effective valley network computed at the Fermi level (in other words, the network of valley lines for which $u < 1/E_F$) must allow one subregion to percolate through the entire system}. In 2D, if a subregion crosses the entire domain then its effective valley network has to be disconnected. The percolation transition of the subregion thus exactly corresponds to the percolation transition of the effective valley network itself.

This theory also allows us to understand how temperature can modify the conductance of the system. On one hand, its increases the amplitude of disorder that enters the Schr\"{o}dinger equation. This tends to localize the quantum states and to reduce the conductance. On the other hand, it explores excited states at higher energies, which are less localized, and it couples the states, allowing electrons to hop between states. The existence of a transition at finite temperature in a macroscopic system therefore depends on the outcome of the competition between these effects.

Finally one should recall that this theory is formally valid in any dimension. In particular, it means that the localization or delocalization properties of a system of $N$ correlated electrons can also be analyzed, in theory, through the knowledge of the statistical properties of the valley network for the landscape $\displaystyle u = \left(-\sum_{i=1}^N \Delta_i + V\right)^{-1} \mathbf{1}$ in a $3N$-dimensional space.

\paragraph*{Conclusion}

What emerges here is a new picture of localization. There are, in fact, not \emph{three} but only \emph{two} types of localization: \emph{low frequency} localization, described by the landscape theory developed in this paper, and \emph{scar} or \emph{high frequency} localization, predicted by the stable orbits of the domain.

Our findings demonstrate that low frequency localization is a universal phenomenon, observed for any type of vibration governed by a spatial differential operator $L$ that derives from an energy form. The geometry of the domain and the properties of the operator interplay to crea\-te a landscape $u$ which entirely determines the localization properties of the system. First, its network of valley lines (in 2D) or surfaces (in 3D) creates a partition of the initial domain into disjoint subregions which shape the spatial distributions of the vibrational modes and identify precisely the regions confining vibrations. Second, the depth of these valleys determines the strength of the confinement within each subregion. The localization of a given mode of eigenvalue~$\lambda$ is controlled by the effective valley network defined as the subset of the entire original valley network subject to the condition $u<1/\lambda$. Consequently, the relative number of localized modes decreases at high frequencies. After partitioning by the landscape, a complex vibrational system can be understood as a collection of weakly coupled subregions whose coupling increases with frequency. 


The theory holds for systems of irregular geometry as well as for disordered ones. In this framework, \emph{Anderson localization} arises as a specific form of weak localization, ``strengthened" by the extremely rough landscape generated by a random potential. More generally, the macroscopic properties materials or systems in which localization plays an essential role, can now be reformulated from the geometrical and analytical characteristics of the effective valley network.


This theory of localization opens a number of new problems. Let us mention a few: In the case of domain with fractal boundary, can one relate the asymptotic distribution of eigenmodes with the scaling properties of the valley network? Can one deduce the thermodynamical behavior of non interacting bosons or fermions in a disordered system from the knowledge of its valley network? What are the statistical properties of the landscape of $N$ interacting particles?

Finally, one should stress that the effective valley network is promising to become a new tool of primary importance for designing systems with specific vibrational properties. To this end, future studies should investigate in detail the relationship between the geometry of a system (irregular or fractal), the characteristics of the wave operator (order, non-homogeneity, possibly stochastic), and the properties of one key mapping, the resulting valley network.

\begin{scilastnote}
\item Part of this work was completed during the visit of the second author to the ENS Cachan. Both authors were partially supported by the ENS Cachan through the Farman program. The first author is also partially supported by the ANR Program ``Silent Wall" ANR-06-MAPR-00-18. The second author is partially supported by the Alfred P. Sloan Fellowship, the NSF CAREER Award DMS 1056004, and the US NSF Grant DMS 0758500. The authors also wish to thank Guy David for fruitful discussions.
\end{scilastnote}

\bibliography{scibib}

\bibliographystyle{Science}




\newpage
\paragraph*{SUPPORTING ONLINE MATERIAL}

\paragraph*{Materials and methods}

\paragraph*{Preliminaries}\label{s1}
\setcounter{equation}{0}

Let $\Omega$ be an arbitrary bounded open set in $\RR^n$ and let $L$ be any elliptic differential operator associated to a symmetric positive bilinear form $B$ (an energy integral). Essentially all elliptic operators governing wave propagation, whether in acoustics, mechanics, electromagnetism, or quantum physics, are associated with an energy and fall into this category. The most prominent examples include:
\begin{eqnarray}\label{eq0.1}
\mbox{the Laplacian} \quad &L=-\Delta, \quad &B[u,v]=\int_\Omega\nabla u\,\nabla v \,dx, \\[4pt]
\label{eq0.2} \mbox{the Hamiltonian} \quad &L=-\Delta+V(x), \quad 0\leq V(x)\leq C, \quad &B[u,v]=\int_\Omega\nabla u\,\nabla v + V u v \,dx, \\
\label{eq0.3} \mbox{the bilaplacian} \quad &L=\Delta^2=-\Delta(-\Delta), \quad &B[u,v]=\int_\Omega\Delta u\,\Delta v\,dx,
\end{eqnarray}

\noindent and finally, any second order divergence form elliptic operator 
\begin{equation}\label{eq0}
L=-{\rm div}\,A(x) \nabla, \qquad  B[u,v]=\int_{\Omega} A(x) \nabla u \nabla v\,dx,\end{equation}

\noindent where $A$ is an elliptic real symmetric $n\times n$ matrix with bounded measurable coefficients, that is, 
\begin{equation}\label{eq0.5}
A(x)=\{a_{ij}(x)\}_{i,j=1}^n, \,x\in\Omega, \qquad a_{ij}\in L^\infty(\Omega), \qquad \sum_{i,j=1}^{n} a_{ij}(x) \xi_i\xi_j\geq \lambda |\xi|^2,\,\,\forall\,\xi\in\RR^n,
\end{equation}

\noindent for some $\lambda>0$, and $a_{ij}=a_{ji}, \,\,\forall i,j=1,...,n.$ 

In general,  $L$ is a differential operator of order $2m$, $m\in \NN$, defined in the weak sense: 
\begin{equation}\label{eq0.4}
\int_\Omega Lu \,v\,dx:= B[u,v],\qquad \mbox{for every}\qquad u,v\in \ring H^m(\Omega),
\end{equation}

\noindent where $B$ is a bounded positive bilinear form and $\ring H^m(\Omega)$ is the Sobolev space of functions given by the completion of $C_0^\infty(\Omega)$ in the norm 
\begin{equation}\label{eq2}
\|u\|_{\ring H^m(\Omega)}:=\|\nabla^m u\|_{L^2(\Omega)},
\end{equation}

\noindent and $\nabla^m u$ denotes the vector of all partial derivatives of $u$ of order $m$.
Recall that the Lax-Milgram Lemma ascertains that for every $f\in (\ring H^m(\Omega))^*=:H^{-m}(\Omega)$ the boundary value problem 
\begin{equation}\label{eq3}
Lu=f, \quad u\in \ring H^m(\Omega),
\end{equation}

\noindent has a unique solution understood in the weak sense. 

\noindent {\bf Remark}.   
The weak solution formalism is necessary to treat the Dirichlet boundary problem on an {\it arbitrary} bounded domain. When the boundary, the data, and the coefficients of the equation are sufficiently smooth, the weak solution coincides with the {\it classical solution} and \eqref{eq3} can be written for a second order operator as 
\begin{equation}\label{eq3.1}
Lu=f, \qquad u|_{\po}=0,
\end{equation}

\noindent where $u|_{\po}$ is the usual pointwise limit at the boundary. For a $2m$-th order operator the derivatives up to the order $m-1$ must vanish as well, e.g.,  
\begin{equation}\label{eq3.2}
\Delta^2 u=f, \qquad u|_{\po}=0,\qquad \partial_\nu u|_{\po}=0,
\end{equation}
\noindent where $\partial_\nu$ stays for normal derivative at the boundary.
In any context, the condition $u\in \ring H^m(\Omega)$ automatically prescribes zero Dirichlet boundary data. On rough domains the definitions akin to \eqref{eq3.1}, \eqref{eq3.2} might not make sense, i.e., a pointwise boundary limit might not exist (the solution might be discontinuous at the boundary), and then the Dirichlet data can only be interpreted in the sense of \eqref{eq3}. 
 
The classification of domains in which all solutions to the Laplace's equation are continuous up to the boundary is available due to the celebrated 1924 Wiener criterion \cite{Wiener}. Over the years, Wiener test has been extended to a variety of operators. We shall not concentrate on this issue, let us just mention the results covering all divergence form second order elliptic operators \cite{LSW}, and the bilaplacian in dimension three \cite{Mayboroda2009}. 
 
Here we shall impose no additional restriction on $\po$ or on the coefficients and work in the general context of weak solutions.

\vskip 0.08 in

For later reference, we also define the Green function of $L$, as conventionally, by 
\begin{equation}\label{eq4.1}
L_xG(x,y)=\delta_y(x), \quad \mbox{for all} \,x,y\in\Omega, \quad G(\cdot,y)\in \ring H^{m}(\Omega)
\quad \mbox{for all} \,y\in\Omega,
\end{equation}
\noindent in the sense of \eqref{eq3}, so that 
\begin{equation}\label{eq4.2}
\int_{\RR^n}L_xG(x,y) v(x)\,dx=v(y), \quad y\in \Omega, \end{equation}
\noindent for every $v\in \ring H^{m}(\Omega)$. It is not difficult to show that for a self-adjoint elliptic operator the Green function is symmetric, i.e., $G(x,y)=G(y,x)$, $x,y\in\Omega$.

\paragraph*{Control of the eigenfunctions by the solution to the Dirichlet problem: the landscape}

Let us now turn to the discussion of the eigenfunctions of $L$. Unless otherwise stated,  we assume that $L$ is an elliptic operator in the weak sense described above and that the underlying bilinear form is symmetric, i.e., that $L$ is self-adjoint. 

The Fredholm theory provides a framework  to consider the eigenvalue problem: 
\begin{equation}\label{eq8}
L\varphi=\lambda \varphi, \quad \varphi\in \ring H^{m}(\Omega),
\end{equation}

\noindent where $\lambda\in\RR$. If there exists a non-trivial solution to \eqref{eq8}, interpreted, as before, in the weak sense then the corresponding $\lambda\in\RR$ is called an eigenvalue and $\varphi\in \ring H^{m}(\Omega)$ is an eigenvector. 

\begin{proposition}
\label{p1}Let $\Omega$ be an arbitrary bounded open set, $L$ be a self-adjoint elliptic operator on $\Omega$, and assume that $\varphi\in \ring H^{m}(\Omega)$ is an eigenfunction of $L$ and $\lambda$ is the corresponding eigenvalue, i.e., \eqref{eq8} is satisfied. Then 
\begin{equation}\label{eq10_sup}
\frac{|\varphi(x)|}{\|\varphi\|_{L^{\infty}(\Omega)}}\leq \lambda u(x), \quad\mbox{for all}\,x\in\Omega,\end{equation} 

\noindent provided that $\varphi \in L^\infty(\Omega)$, with
\begin{equation}\label{eq9_sup}
u(x)= \int_\Omega |G(x,y)|\,dy, \qquad x\in\Omega.\end{equation} 

If, in addition, the Green function is non-negative (in the sense of distributions), then 
 $u$ is the solution of the boundary problem  
\begin{equation}\label{eq11_sup}
Lu=1, \quad u\in \ring H^m(\Omega).
\end{equation}
\end{proposition}

\noindent {\bf Remark.} The Green function is positive and eigenfunctions are bounded for the Laplacian \eqref{eq0.1}, the Hamiltonian \eqref{eq0.2}, all second order elliptic operators \eqref{eq0} in all dimensions due to the strong maximum principle (see, e.g., \cite{Gilbarg2001}, Section~8.7). Hence, for all such operators \eqref{eq10_sup}, \eqref{eq11_sup} are valid. 

The situation for the higher order PDEs is more subtle. In fact, even for the bilaplacian the positivity in general fails, and then one has to operate directly with \eqref{eq9_sup}.

\vskip 0.08in
\bp By \eqref{eq8} and \eqref{eq4.1} (with the roles of $x$ and $y$ interchanged) and self-adjointness of $L$, for every $x\in \Omega$
\begin{equation}\label{eq12}
\varphi(x) = \int_\Omega\varphi(y)\,  L_y  G(x,y)\,dy= \int_\Omega L_y \varphi(y)\, G(x,y)\,dy=  \int_\Omega \lambda \,\varphi(y)\, G(x,y)\,dy,
\end{equation}

\noindent and hence, 
\begin{equation}\label{eq13}
|\varphi(x)|\leq \lambda \,\|\varphi\|_{L^\infty(\Omega)}\int_{\Omega}|G(x,y)|\,dy, \quad x\in \Omega,
\end{equation}

\noindent as desired. Moreover, if the Green function is positive, 
\begin{equation}\label{eq14}
\int_{\Omega}|G(x,y)|\,dy=\int_{\Omega}G(x,y)\cdot 1\,dy, \quad x\in \Omega,
\end{equation}

\noindent which is by definition a solution of \eqref{eq11_sup}.\ep

Vaguely speaking, the inequality \eqref{eq10_sup} provides the ``landscape of localization", as the map of $u$ in \eqref{eq9_sup} -- \eqref{eq11_sup} draws lines separating subdomains which will ``host" localized eigenmodes. We now discuss in details the situation on these subdomains.

\paragraph*{Analysis of localized modes on the subdomains}

The gist of the forthcoming discussion is that, roughly speaking, a mode of $\Omega$ localized to a subdomain $D\subset \Omega$ must be fairly close to an eigenmode of this subdomain, and an eigenvalue of $\Omega$ for which localization takes place, must be close to some eigenvalue of $D$.

To this end, let $\varphi$ be one of the eigenmodes of $\Omega$, which exhibits localization to $D$, a subdomain of $\Omega$. This means, in particular, that the boundary values of $\varphi$ on $\partial D$ are small. In fact, in the present context the correct way to interpret ``smallness" of $\varphi$ on the boundary of $D$ is in terms of the smallness of an  $L$-harmonic function, with the same data as $\varphi$ on $\partial D$. 

To be rigorous,  let us define $\eps=\eps_{\varphi}>0$ as
\begin{eqnarray}\label{eq14.1}\nonumber
&& \mbox{$\eps=\|v\|_{L^2(D)}$, where $v\in H^m(D)$ is such that} \\
&& \mbox{$w:=\varphi-v\in \ring H^m(D)$ (that is, the boundary values of $\varphi$ and $v$ on $\partial D$ coincide),} \\ && \mbox{and $Lv=0$ on $D$ in the sense of distributions.}\nonumber
\end{eqnarray}

Then the following Proposition holds.

\begin{proposition}\label{pRes} Let $\Omega$ be an arbitrary bounded open set, $L$ be a self-adjoint elliptic operator on $\Omega$, and  $\varphi\in \ring H^{m}(\Omega)$ be an eigenmode of $L$. Suppose further that $D$ is a subset of $\Omega$ and denote by $\eps$ the norm of the boundary data of $\varphi$ on $\partial D$ in the sense of \eqref{eq14.1}.

Denote by $\lambda$ the eigenvalue corresponding to $\varphi$. Then either $\lambda$ is an eigenvalue of $L$ in $D$ or 
\begin{equation}\label{eq19_sup}
\|\varphi\|_{L^2(D)}  \leq  \left(1+\max_{\lambda_k(D)} \left\{\left|1-\frac{\lambda_k(D)}{\lambda}\right|^{-1}\right\}\right) \eps,
\end{equation}

\noindent where the maximum is taken over all eigenvalues of $L$ in $D$.

\end{proposition}

\bp First of all, note that \eqref{eq14.1} implies
\begin{equation}\label{eq15}
(L-\lambda)w=\lambda v \quad\mbox{on } D,
\end{equation}

\noindent as usually, in the sense of distributions. If $\lambda$ is an eigenvalue of $D$, there is nothing to prove. If $\lambda$ is not an eigenvalue of $D$, we claim that 
\begin{equation}\label{eq16}
\|w\|_{L^2(D)}  \leq \max_{\lambda_k(D)} \left\{\frac{1}{|\lambda-\lambda_k(D)|}\right\} \|\lambda v\|_{L^2(D)}.
\end{equation}

Indeed, in our setup, the eigenvalues of $L$ are real, positive, at most countable,  and moreover, there exists an orthonormal basis of $L^2(D)$  formed by the eigenfunctions of $L$ on $D$, $\{\psi_{k, D}\}_{k}$. In particular, for every $f\in \ring H^m(D)\subset L^2(D)$ we can write
\begin{equation}\label{eq16.1}
f=\sum _{k} c_k(f) \psi_k, \qquad c_k(f)=\int_D f\,\psi_k\,dx,  
\end{equation}

\noindent with the convergence in $L^2(D)$, and $\|f\|_{L^2(D)}=\left(\sum _{k}c_k(f)^2\right)^{1/2}$. Moreover, such a  series $\sum _{k} c_k(f) \psi_k$ converges in $\ring H^m(D)$ as well and $\{\psi_{k, D}\}_k$ form an orthogonal basis of $\ring H^m(D)$. These considerations follow from ellipticity and self-adjointness of $L$ in a standard way using the machinery of functional analysis (see, e.g., \cite{Evans2010}, pp. 355--358 treating the case of the second order operator of the type \eqref{eq0}).

Therefore, for every $\lambda$ not belonging to the spectrum of $L$ on $D$ and $w\in \ring H^m(D)$ with $(L-\lambda) w \in L^2(D)$ (cf. \eqref{eq15}) we have 
\begin{equation}\label{eq16.1.0}
 \|(L-\lambda)w\|_{L^2(D)} = \left\|\sum_k c_k((L-\lambda)w)\psi_k\right \|_{L^2(D)},
\end{equation}

\noindent where 
\begin{eqnarray}\label{eq16.1.0.1}
c_k((L-\lambda)w)&=&\int_D(L-\lambda)w \,\psi_k\,dx= \int_Dw \,(L-\lambda) \psi_k\,dx\\[4pt]
&=& (\lambda_k(D)-\lambda)  \int_Dw \, \psi_k\,dx=(\lambda_k(D)-\lambda) c_k(w).
\end{eqnarray}

\noindent Hence,
\begin{eqnarray}\label{eq16.2}\nonumber
 \|(L-\lambda)w\|_{L^2(D)} &=& \left\|\sum_k (\lambda_k(D)-\lambda) c_k(w)\psi_k\right \|_{L^2(D)}=\left(\sum_k (\lambda_k(D)-\lambda)^2 c_k(w)^2\right)^{1/2}\\[4pt]\nonumber
 &\geq& \min_{\lambda_k(D)} |\lambda_k(D)-\lambda| \left(\sum_k c_k(w)^2\right)^{1/2}= \min_{\lambda_k(D)}  |\lambda_k(D)-\lambda| \,\|w\|_{L^2(D)}, 
\end{eqnarray}

\noindent which leads to  inequality \eqref{eq16}.

Going further, \eqref{eq16} yields
\begin{equation}\label{eq17}
\|w\|_{L^2(D)}  \leq \max_{\lambda_k(D)} \left\{\left|1-\frac{\lambda_k(D)}{\lambda}\right|^{-1}\right\} \|v\|_{L^2(D)}\leq  \max_{\lambda_k(D)} \left\{\left|1-\frac{\lambda_k(D)}{\lambda}\right|^{-1}\right\} \eps,
\end{equation}

\noindent and therefore, 
\begin{equation}\label{eq18}
\|\varphi\|_{L^2(D)}  \leq  \left(1+\max_{\lambda_k(D)} \left\{\left|1-\frac{\lambda_k(D)}{\lambda}\right|^{-1}\right\}\right) \eps,
\end{equation}

\noindent as desired.

\end{document}